# Quadratic recurrence equations - exact explicit solution of period four fixed points functions in bifurcation diagram

**Gvozden Rukavina**

Jurjenici 80, 51215 Kastav, Croatia
e-mail: gvozden.rukavina@ri.t-com.hr

The article presents exact solution of quadratic recurrence final state functions for cycle of period 4. The solution is demonstrated for the quadratic map and the logistic map and the explanation of the solution is given in the last part of the article.

## 1. Fixed points functions of period 4 for the quadratic map

Recurrence equation:

$$z_{n+1} = z_n^2 + c$$

Time sequence gives the following relations:

$$z_1 = z_1$$

$$z_2 = z_1^2 + c$$

$$z_3 = z_2^2 + c$$

$$z_4 = z_3^2 + c$$

$$z_5 = z_4^2 + c = z_1$$

For convenience, the fixed points functions are not indexed by time sequence but to follow the spatial order of bifurcation branches (starting from the lowest one). In order to make it easier to follow, the appropriate values of functions corresponding to supper attractive points at $c$ = -1.31070 are indicated beside them. The respective relations are as follows:

$$z_{41} = z_1 = z_{43}^2 + c \qquad\qquad -1.31070$$

$$z_{42} = z_3 = z_{44}^2 + c \qquad\qquad -1.14486$$

$$z_{43} = z_4 = z_{42}^2 + c \qquad\qquad 0.0$$

$$z_{44} = z_2 = z_{41}^2 + c \qquad\qquad 0.407239$$

The above functions are determined by the following expressions:

$$z_{41} = z_{m1} - \sqrt{-c - z_{m1}^2 + z_{m2}}$$

$$z_{42} = z_{m1} + \sqrt{-c - z_{m1}^2 + z_{m2}}$$

$$z_{43} = z_{m2} - \sqrt{-c - z_{m2}^2 + z_{m1}}$$

$$z_{44} = z_{m2} + \sqrt{-c - z_{m2}^2 + z_{m1}}$$

where:

$$z_{m1} = \frac{1}{2}(z_{41} + z_{42})$$

$$z_{m2} = \frac{1}{2}(z_{43} + z_{44})$$

are the mean values of lower / upper bifurcation pitchfork respectively. This mean value functions are determined by the following relations:

$$z_{m1} = Z_m - \sqrt{Z_m^2 + \frac{1}{4}}$$

$$z_{m2} = Z_m + \sqrt{Z_m^2 + \frac{1}{4}}$$

where $Z_m$ presents the mean value of all four fixed points functions i.e.:

$$Z_m = \frac{1}{2}(z_{m1} + z_{m2}) = \frac{1}{4}(z_{41} + z_{42} + z_{43} + z_{44})$$

Finally, $Z_m$ is the pure function of $c$ and presents a root of a simple cubic polynomial equation as is shown in the third part of the article:

$$Z_m = -\frac{1}{2}\left(\left(\frac{1}{4} - \sqrt{\frac{1}{16} - \left(-\frac{1}{4} - \frac{c}{3}\right)^3}\right)^{1/3} + \left(\frac{1}{4} + \sqrt{\frac{1}{16} - \left(-\frac{1}{4} - \frac{c}{3}\right)^3}\right)^{1/3}\right)$$

By applying the above defined substitutions, the fixed points functions are determined as follows:

$$z_{41} = Z_m - \sqrt{\frac{1}{4} + Z_m^2} - \sqrt{-c + Z_m + \sqrt{\frac{1}{4} + Z_m^2} - \left(Z_m - \sqrt{\frac{1}{4} + Z_m^2}\right)^2}$$

$$z_{42} = Z_m - \sqrt{\frac{1}{4} + Z_m^2} + \sqrt{-c + Z_m + \sqrt{\frac{1}{4} + Z_m^2} - \left(Z_m - \sqrt{\frac{1}{4} + Z_m^2}\right)^2}$$

$$z_{43} = Z_m + \sqrt{\frac{1}{4} + Z_m^2} - \sqrt{-c + Z_m - \sqrt{\frac{1}{4} + Z_m^2} - \left(Z_m + \sqrt{\frac{1}{4} + Z_m^2}\right)^2}$$

$$z_{44} = Z_m + \sqrt{\frac{1}{4} + Z_m^2} + \sqrt{-c + Z_m - \sqrt{\frac{1}{4} + Z_m^2} - \left(Z_m + \sqrt{\frac{1}{4} + Z_m^2}\right)^2}$$

or in the fully developed form:

$$z_{41} = -\sqrt{\frac{1}{4} + \frac{1}{4}\left(-\left(\frac{1}{4} - \sqrt{\frac{1}{16} - \left(-\frac{1}{4} - \frac{c}{3}\right)^3}\right)^{1/3} - \left(\frac{1}{4} + \sqrt{\frac{1}{16} - \left(-\frac{1}{4} - \frac{c}{3}\right)^3}\right)^{1/3}\right)^2} +$$

$$\frac{1}{2}\left(-\left(\frac{1}{4} - \sqrt{\frac{1}{16} - \left(-\frac{1}{4} - \frac{c}{3}\right)^3}\right)^{1/3} - \left(\frac{1}{4} + \sqrt{\frac{1}{16} - \left(-\frac{1}{4} - \frac{c}{3}\right)^3}\right)^{1/3}\right) -$$

$$\sqrt{\left(-\left(-\sqrt{\frac{1}{4} + \frac{1}{4}\left(-\left(\frac{1}{4} - \sqrt{\frac{1}{16} - \left(-\frac{1}{4} - \frac{c}{3}\right)^3}\right)^{1/3} - \left(\frac{1}{4} + \sqrt{\frac{1}{16} - \left(-\frac{1}{4} - \frac{c}{3}\right)^3}\right)^{1/3}\right)^2} + \right.\right.}$$

$$\frac{1}{2}\left(-\left(\frac{1}{4} - \sqrt{\frac{1}{16} - \left(-\frac{1}{4} - \frac{c}{3}\right)^3}\right)^{1/3} - \left(\frac{1}{4} + \sqrt{\frac{1}{16} - \left(-\frac{1}{4} - \frac{c}{3}\right)^3}\right)^{1/3}\right)\Bigg)^2 +$$

$$\sqrt{\frac{1}{4} + \frac{1}{4}\left(-\left(\frac{1}{4} - \sqrt{\frac{1}{16} - \left(-\frac{1}{4} - \frac{c}{3}\right)^3}\right)^{1/3} - \left(\frac{1}{4} + \sqrt{\frac{1}{16} - \left(-\frac{1}{4} - \frac{c}{3}\right)^3}\right)^{1/3}\right)^2} +$$

$$\left.\frac{1}{2}\left(-\left(\frac{1}{4} - \sqrt{\frac{1}{16} - \left(-\frac{1}{4} - \frac{c}{3}\right)^3}\right)^{1/3} - \left(\frac{1}{4} + \sqrt{\frac{1}{16} - \left(-\frac{1}{4} - \frac{c}{3}\right)^3}\right)^{1/3}\right) - c\right)$$

$$z_{42} = -\sqrt{\frac{1}{4} + \frac{1}{4}\left(-\left(\frac{1}{4} - \sqrt{\frac{1}{16} - \left(-\frac{1}{4} - \frac{c}{3}\right)^3}\right)^{1/3} - \left(\frac{1}{4} + \sqrt{\frac{1}{16} - \left(-\frac{1}{4} - \frac{c}{3}\right)^3}\right)^{1/3}\right)^2} +$$

$$\frac{1}{2}\left(-\left(\frac{1}{4} - \sqrt{\frac{1}{16} - \left(-\frac{1}{4} - \frac{c}{3}\right)^3}\right)^{1/3} - \left(\frac{1}{4} + \sqrt{\frac{1}{16} - \left(-\frac{1}{4} - \frac{c}{3}\right)^3}\right)^{1/3}\right) +$$

$$\sqrt{\left[-\left(-\sqrt{\frac{1}{4} + \frac{1}{4}\left(-\left(\frac{1}{4} - \sqrt{\frac{1}{16} - \left(-\frac{1}{4} - \frac{c}{3}\right)^3}\right)^{1/3} - \left(\frac{1}{4} + \sqrt{\frac{1}{16} - \left(-\frac{1}{4} - \frac{c}{3}\right)^3}\right)^{1/3}\right)^2}\right)\right.} +$$

$$\left.\frac{1}{2}\left(-\left(\frac{1}{4} - \sqrt{\frac{1}{16} - \left(-\frac{1}{4} - \frac{c}{3}\right)^3}\right)^{1/3} - \left(\frac{1}{4} + \sqrt{\frac{1}{16} - \left(-\frac{1}{4} - \frac{c}{3}\right)^3}\right)^{1/3}\right)\right]^2 +$$

$$\sqrt{\frac{1}{4} + \frac{1}{4}\left(-\left(\frac{1}{4} - \sqrt{\frac{1}{16} - \left(-\frac{1}{4} - \frac{c}{3}\right)^3}\right)^{1/3} - \left(\frac{1}{4} + \sqrt{\frac{1}{16} - \left(-\frac{1}{4} - \frac{c}{3}\right)^3}\right)^{1/3}\right)^2} +$$

$$\frac{1}{2}\left(-\left(\frac{1}{4} - \sqrt{\frac{1}{16} - \left(-\frac{1}{4} - \frac{c}{3}\right)^3}\right)^{1/3} - \left(\frac{1}{4} + \sqrt{\frac{1}{16} - \left(-\frac{1}{4} - \frac{c}{3}\right)^3}\right)^{1/3}\right) - c$$

$$z_{43} = \sqrt{\frac{1}{4} + \frac{1}{4}\left(-\left(\frac{1}{4} - \sqrt{\frac{1}{16} - \left(-\frac{1}{4} - \frac{c}{3}\right)^3}\right)^{1/3} - \left(\frac{1}{4} + \sqrt{\frac{1}{16} - \left(-\frac{1}{4} - \frac{c}{3}\right)^3}\right)^{1/3}\right)^2} +$$

$$\frac{1}{2}\left(-\left(\frac{1}{4} - \sqrt{\frac{1}{16} - \left(-\frac{1}{4} - \frac{c}{3}\right)^3}\right)^{1/3} - \left(\frac{1}{4} + \sqrt{\frac{1}{16} - \left(-\frac{1}{4} - \frac{c}{3}\right)^3}\right)^{1/3}\right) -$$

$$\sqrt{\left[-\left(\sqrt{\frac{1}{4} + \frac{1}{4}\left(-\left(\frac{1}{4} - \sqrt{\frac{1}{16} - \left(-\frac{1}{4} - \frac{c}{3}\right)^3}\right)^{1/3} - \left(\frac{1}{4} + \sqrt{\frac{1}{16} - \left(-\frac{1}{4} - \frac{c}{3}\right)^3}\right)^{1/3}\right)^2}\right)\right.} +$$

$$\left.\frac{1}{2}\left(-\left(\frac{1}{4} - \sqrt{\frac{1}{16} - \left(-\frac{1}{4} - \frac{c}{3}\right)^3}\right)^{1/3} - \left(\frac{1}{4} + \sqrt{\frac{1}{16} - \left(-\frac{1}{4} - \frac{c}{3}\right)^3}\right)^{1/3}\right)\right]^2 -$$

$$\sqrt{\frac{1}{4} + \frac{1}{4}\left(-\left(\frac{1}{4} - \sqrt{\frac{1}{16} - \left(-\frac{1}{4} - \frac{c}{3}\right)^3}\right)^{1/3} - \left(\frac{1}{4} + \sqrt{\frac{1}{16} - \left(-\frac{1}{4} - \frac{c}{3}\right)^3}\right)^{1/3}\right)^2} +$$

$$\frac{1}{2}\left(-\left(\frac{1}{4} - \sqrt{\frac{1}{16} - \left(-\frac{1}{4} - \frac{c}{3}\right)^3}\right)^{1/3} - \left(\frac{1}{4} + \sqrt{\frac{1}{16} - \left(-\frac{1}{4} - \frac{c}{3}\right)^3}\right)^{1/3}\right) - c$$

$$z_{44} = \sqrt{\frac{1}{4} + \frac{1}{4}\left(-\left(\frac{1}{4} - \sqrt{\frac{1}{16} - \left(-\frac{1}{4} - \frac{c}{3}\right)^3}\right)^{1/3} - \left(\frac{1}{4} + \sqrt{\frac{1}{16} - \left(-\frac{1}{4} - \frac{c}{3}\right)^3}\right)^{1/3}\right)^2} +$$

$$\frac{1}{2}\left(-\left(\frac{1}{4} - \sqrt{\frac{1}{16} - \left(-\frac{1}{4} - \frac{c}{3}\right)^3}\right)^{1/3} - \left(\frac{1}{4} + \sqrt{\frac{1}{16} - \left(-\frac{1}{4} - \frac{c}{3}\right)^3}\right)^{1/3}\right) +$$

$$\sqrt{\left(-\left(\sqrt{\frac{1}{4} + \frac{1}{4}\left(-\left(\frac{1}{4} - \sqrt{\frac{1}{16} - \left(-\frac{1}{4} - \frac{c}{3}\right)^3}\right)^{1/3} - \left(\frac{1}{4} + \sqrt{\frac{1}{16} - \left(-\frac{1}{4} - \frac{c}{3}\right)^3}\right)^{1/3}\right)^2} + \right.\right.}$$

$$\left.\left. \frac{1}{2}\left(-\left(\frac{1}{4} - \sqrt{\frac{1}{16} - \left(-\frac{1}{4} - \frac{c}{3}\right)^3}\right)^{1/3} - \left(\frac{1}{4} + \sqrt{\frac{1}{16} - \left(-\frac{1}{4} - \frac{c}{3}\right)^3}\right)^{1/3}\right)\right)^2 -\right.$$

$$\sqrt{\frac{1}{4} + \frac{1}{4}\left(-\left(\frac{1}{4} - \sqrt{\frac{1}{16} - \left(-\frac{1}{4} - \frac{c}{3}\right)^3}\right)^{1/3} - \left(\frac{1}{4} + \sqrt{\frac{1}{16} - \left(-\frac{1}{4} - \frac{c}{3}\right)^3}\right)^{1/3}\right)^2} +$$

$$\left.\frac{1}{2}\left(-\left(\frac{1}{4} - \sqrt{\frac{1}{16} - \left(-\frac{1}{4} - \frac{c}{3}\right)^3}\right)^{1/3} - \left(\frac{1}{4} + \sqrt{\frac{1}{16} - \left(-\frac{1}{4} - \frac{c}{3}\right)^3}\right)^{1/3}\right) - c\right)$$

Here below is presented the picture of bifurcation diagram with plotted fixed points functions for cycles with period 1, 2 and 4. The functions are plotted also in the area behind the bifurcation point because they also exist there, only changing the attraction quality from stable to unstable. They are passing through the fractal analogs of band merging point as well as through many other characteristic knots of iterative functions as is presented in the third part of the article.

The one and two period cycles are presented by well known expressions:

$$z_1 = \frac{1}{2}\left(1 - \sqrt{1 - 4c}\right)$$

$$z_{21} = \frac{1}{2}\left(-1 - \sqrt{-3 - 4c}\right)$$

$$z_{22} = \frac{1}{2}\left(-1 + \sqrt{-3 - 4c}\right)$$

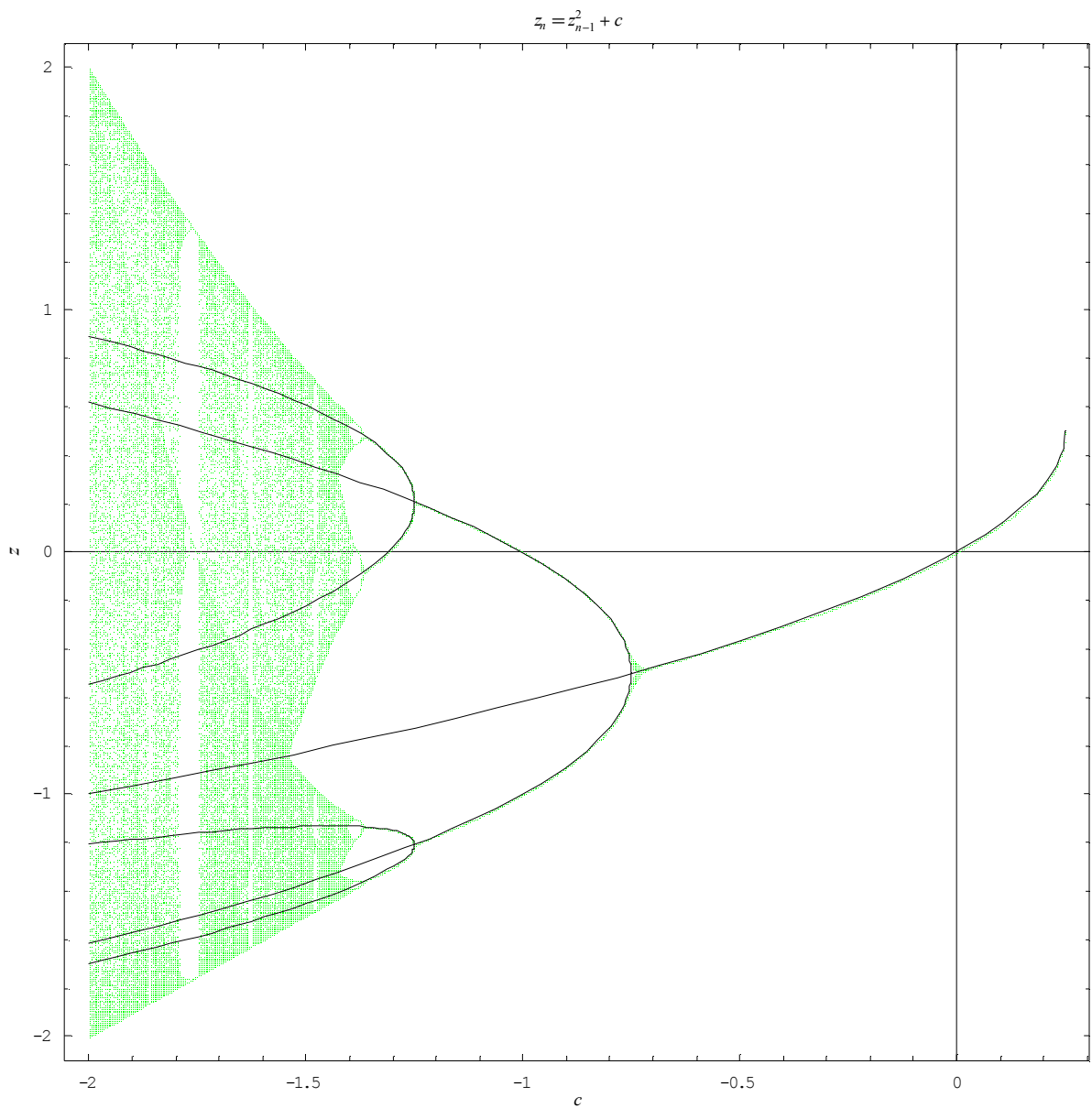

$$z_n = z_{n-1}^2 + c$$

## 2. Fixed points functions of period 4 for the logistic map

The solution of fixed points functions presented above in part 1. is valid also for other quadratic iterators by application of appropriate transformation of coordinates. The transformation is not elaborated because it is very simple and only the results are presented. The meaning of indexes is equivalent to the principle described above except that, due to the transformation rule, the analogue index numbers of fixed points functions are increasing from the highest value function downwards.

Recurrence equation:

$$x_{n+1} = a * x (1 - x)$$

Time sequence relations:

$$x_1 = x_1$$

$$x_2 = a * x_1 (1 - x_1)$$

$$x_3 = a * x_2 (1 - x_2)$$

$$x_4 = a * x_3 (1 - x_3)$$

$$x_5 = a * x_4 (1 - x_4) = x_1$$

The following expressions and relations are defined, with identification by bifurcation branches order (starting from the highest one):

fixed points functions

$$x_{41} = x_1 = a * x_{43} (1 - x_{43})$$

$$x_{42} = x_3 = a * x_{44} (1 - x_{44})$$

$$x_{43} = x_4 = a * x_{42} (1 - x_{42})$$

$$x_{44} = x_2 = a * x_{41} (1 - x_{41})$$

mean values of respective fixed points functions pairs

$$x_{m1} = \frac{1}{2} (x_{41} + x_{42})$$

$$x_{m2} = \frac{1}{2} (x_{43} + x_{44})$$

mean value of all four fixed points functions

$$X_m = \frac{1}{2} (x_{m1} + x_{m2}) = \frac{1}{4} (x_{41} + x_{42} + x_{43} + x_{44})$$

The expressions determining the above functions are provided from the solutions for the quadratic map, given in the first part of the article, by the following transformation relations:

$$c = \frac{1}{4} (2 - a) a$$

$$z = a\left(\frac{1}{2} - x\right)$$

Applying the above substitutions and settling the expressions the following results for the above listed functions are yielded:

$$x_{41} = x_{m1} + \sqrt{x_{m1} - x_{m1}^2 - \frac{x_{m2}}{a}}$$

$$x_{42} = x_{m1} - \sqrt{x_{m1} - x_{m1}^2 - \frac{x_{m2}}{a}}$$

$$x_{43} = x_{m2} + \sqrt{-\frac{x_{m1}}{a} + x_{m2} - x_{m2}^2}$$

$$x_{44} = x_{m2} - \sqrt{-\frac{x_{m1}}{a} + x_{m2} - x_{m2}^2}$$

$$x_{m1} = X_m + \sqrt{\frac{1}{4} + \frac{1}{4a^2} - X_m + X_m^2}$$

$$x_{m2} = X_m - \sqrt{\frac{1}{4} + \frac{1}{4a^2} - X_m + X_m^2}$$

$$X_m = \frac{1}{2} + \frac{1}{2a}\left(\left(\frac{1}{4} - \sqrt{\frac{1}{16} - \frac{(-3+a)^3(1+a)^3}{1728}}\right)^{1/3} + \left(\frac{1}{4} + \sqrt{\frac{1}{16} - \frac{(-3+a)^3(1+a)^3}{1728}}\right)^{1/3}\right)$$

Having applied the above relations, the fixed points functions are determined by the following expressions:

$$x_{41} = X_m + \sqrt{\frac{1}{4} + \frac{1}{4a^2} - X_m + X_m^2} +$$

$$\sqrt{X_m + \sqrt{\frac{1}{4} + \frac{1}{4a^2} - X_m + X_m^2} - \frac{X_m - \sqrt{\frac{1}{4} + \frac{1}{4a^2} - X_m + X_m^2}}{a} - \left(X_m + \sqrt{\frac{1}{4} + \frac{1}{4a^2} - X_m + X_m^2}\right)^2}$$

$$x_{42} = X_m + \sqrt{\frac{1}{4} + \frac{1}{4a^2} - X_m + X_m^2} -$$

$$\sqrt{X_m + \sqrt{\frac{1}{4} + \frac{1}{4a^2} - X_m + X_m^2} - \frac{X_m - \sqrt{\frac{1}{4} + \frac{1}{4a^2} - X_m + X_m^2}}{a} - \left(X_m + \sqrt{\frac{1}{4} + \frac{1}{4a^2} - X_m + X_m^2}\right)^2}$$

$$x_{43} = X_m - \sqrt{\frac{1}{4} + \frac{1}{4a^2} - X_m + X_m^2} +$$

$$\sqrt{X_m - \sqrt{\frac{1}{4} + \frac{1}{4a^2} - X_m + X_m^2} - \left(X_m - \sqrt{\frac{1}{4} + \frac{1}{4a^2} - X_m + X_m^2}\right)^2 - \frac{X_m + \sqrt{\frac{1}{4} + \frac{1}{4a^2} - X_m + X_m^2}}{a}}$$

$$x_{44} = X_m - \sqrt{\frac{1}{4} + \frac{1}{4a^2} - X_m + X_m^2} -$$

$$\sqrt{X_m - \sqrt{\frac{1}{4} + \frac{1}{4a^2} - X_m + X_m^2} - \left(X_m - \sqrt{\frac{1}{4} + \frac{1}{4a^2} - X_m + X_m^2}\right)^2 - \frac{X_m + \sqrt{\frac{1}{4} + \frac{1}{4a^2} - X_m + X_m^2}}{a}}$$

As the fully developed form of the expressions is presented in the first part of the article, only one of the four expressions is presented in its fully developed form below and the others can be easily calculated from the functions given above:

$$x_{41} = \frac{1}{2} + \frac{1}{2a}\left(\left(\frac{1}{4} - \sqrt{\frac{1}{16} - \frac{(-3+a)^3(1+a)^3}{1728}}\right)^{1/3} + \left(\frac{1}{4} + \sqrt{\frac{1}{16} - \frac{(-3+a)^3(1+a)^3}{1728}}\right)^{1/3}\right) +$$

$$\frac{1}{2a}\sqrt{\frac{3-2a+a^2}{6} + \left(\frac{1}{4} - \sqrt{\frac{1}{16} - \frac{(-3+a)^3(1+a)^3}{1728}}\right)^{2/3} + \left(\frac{1}{4} + \sqrt{\frac{1}{16} - \frac{(-3+a)^3(1+a)^3}{1728}}\right)^{2/3}} +$$

$$\sqrt{\frac{1}{2} + \frac{1}{2a}\left(\left(\frac{1}{4} - \sqrt{\frac{1}{16} - \frac{(-3+a)^3(1+a)^3}{1728}}\right)^{1/3} + \left(\frac{1}{4} + \sqrt{\frac{1}{16} - \frac{(-3+a)^3(1+a)^3}{1728}}\right)^{1/3}\right) +$$

$$\frac{1}{2a}\sqrt{\frac{3-2a+a^2}{6} + \left(\frac{1}{4} - \sqrt{\frac{1}{16} - \frac{(-3+a)^3(1+a)^3}{1728}}\right)^{2/3} + \left(\frac{1}{4} + \sqrt{\frac{1}{16} - \frac{(-3+a)^3(1+a)^3}{1728}}\right)^{2/3}} -$$

$$\frac{1}{a}\left(\frac{1}{2} + \frac{1}{2a}\left(\left(\frac{1}{4} - \sqrt{\frac{1}{16} - \frac{(-3+a)^3(1+a)^3}{1728}}\right)^{1/3} + \left(\frac{1}{4} + \sqrt{\frac{1}{16} - \frac{(-3+a)^3(1+a)^3}{1728}}\right)^{1/3}\right) - \frac{1}{2a}$$

$$\sqrt{\left(\frac{3-2a+a^2}{6} + \left(\frac{1}{4} - \sqrt{\frac{1}{16} - \frac{(-3+a)^3(1+a)^3}{1728}}\right)^{2/3} + \left(\frac{1}{4} + \sqrt{\frac{1}{16} - \frac{(-3+a)^3(1+a)^3}{1728}}\right)^{2/3}\right)} -$$

$$\left(\frac{1}{2} + \frac{1}{2a}\left(\left(\frac{1}{4} - \sqrt{\frac{1}{16} - \frac{(-3+a)^3(1+a)^3}{1728}}\right)^{1/3} + \left(\frac{1}{4} + \sqrt{\frac{1}{16} - \frac{(-3+a)^3(1+a)^3}{1728}}\right)^{1/3}\right) + \frac{1}{2a}$$

$$\sqrt{\left(\frac{3-2a+a^2}{6} + \left(\frac{1}{4} - \sqrt{\frac{1}{16} - \frac{(-3+a)^3(1+a)^3}{1728}}\right)^{2/3} + \left(\frac{1}{4} + \sqrt{\frac{1}{16} - \frac{(-3+a)^3(1+a)^3}{1728}}\right)^{2/3}\right)^2}}$$

The resulting fixed points functions for cycle with period four are plotted here below over the final state diagram. Beside, the one and two period cycles are also presented by known expressions:

$$x_{11} = \frac{-1+a}{a}$$

$$x_{21} = \frac{1+a+\sqrt{-3-2a+a^2}}{2a}$$

$$x_{22} = \frac{1+a-\sqrt{-3-2a+a^2}}{2a}$$

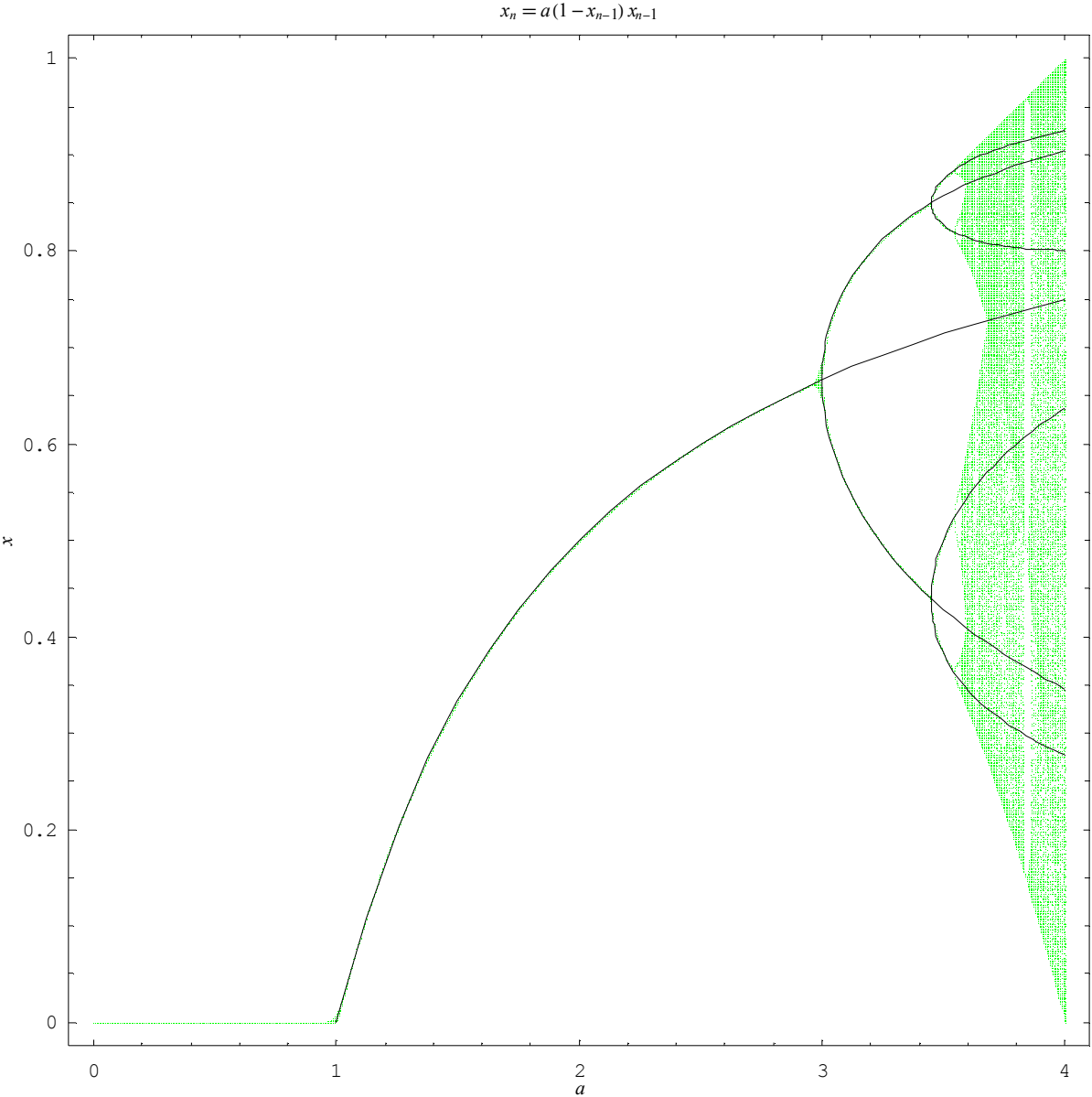

$$x_n = a(1-x_{n-1})x_{n-1}$$

## 3. Fixed points functions of period 4 for the quadratic recurrence equations – explanation of the solution

The procedure of gaining the solution for cycle four fixed points functions is presented here in the way as it was attained and using the recurrence equation which was analysed. That equation is equivalent to the quadratic recurrence equation presented in the first part of the article except that it is mirrored around vertical axes in order to work with positive values of parameter.

Recurrence equation:

$$y_{n+1} = y_n^2 - c$$

Time sequence of recurrence:

$$y_1 = y_1$$

$$y_2 = y_1^2 - c$$

$$y_3 = y_2^2 - c$$

$$y_4 = y_3^2 - c$$

$$y_5 = y_4^2 - c = y_1$$

Identification by bifurcation branches order

$$y_{41} = y_1 = y_{43}^2 - c$$

$$y_{42} = y_3 = y_{44}^2 - c$$

$$y_{43} = y_4 = y_{42}^2 - c$$

$$y_{44} = y_2 = y_{41}^2 - c$$

Following the logic of cycles with period 1 and 2 we may assume the form of the expressions for $y_{4i}$ as follows:

$$y_{41} = y_{m1} - a\sqrt{\varphi}$$

$$y_{42} = y_{m1} + a\sqrt{\varphi}$$

$$y_{43} = y_{m2} - b\sqrt{\varphi}$$

$$y_{44} = y_{m2} + b\sqrt{\varphi}$$

where

$$y_{m1} = \frac{1}{2}(y_{41} + y_{42})$$

$$y_{m2} = \frac{1}{2}(y_{43} + y_{44})$$

represent mean values of respective bifurcation branches.
Parameter $\varphi$ has been assumed to be:

$$\varphi = c - c_{b4}$$

where $c_{b4}$ marks bifurcation 4 onset point.

Having introduced the above assumptions we yield the following results:

$y_1 \rightarrow$

$$y_{41} = y_{m1} - a\sqrt{\varphi}$$

$y_2 \rightarrow$

$$y_{44} = y_{m2} + b\sqrt{\varphi} = y_{41}^2 - c = (y_{m1} - a\sqrt{\varphi})^2 - c =$$
$$= y_{m1}^2 + a^2\varphi - c - 2y_{m1}a\sqrt{\varphi}$$

Resulting relations:

$$y_{m2} = y_{m1}^2 + a^2\varphi - c \qquad (1)$$

$$b = -2y_{m1}a \qquad (2)$$

$y_3 \rightarrow$

$$y_{42} = y_{m1} + a\sqrt{\varphi} = y_{44}^2 - c = (y_{m2} + b\sqrt{\varphi})^2 - c =$$
$$= y_{m2}^2 + b^2\varphi - c + 2y_{m2}b\sqrt{\varphi} =$$
$$= (y_{m2} - 2y_{m1}a\sqrt{\varphi})^2 - c = y_{m2}^2 + 4a^2\varphi y_{m1}^2 - c - 4y_{m1}y_{m2}a\sqrt{\varphi}$$

Resulting relations:

$$y_{m1} = y_{m2}^2 + b^2\varphi - c = y_{m2}^2 + 4y_{m1}^2 a^2\varphi - c \qquad (3)$$

$$a = 2y_{m2}b = -4y_{m1}y_{m2}a \qquad (4)$$

The outcomes from the above results are the following:

(1)→
$$a\sqrt{\varphi} = \sqrt{c - y_{m1}^2 + y_{m2}}$$

(3)→
$$b\sqrt{\varphi} = \sqrt{c - y_{m2}^2 + y_{m1}}$$

This determines the expressions for the fixed points functions as follows:

$$y_{41} = y_{m1} - \sqrt{c - y_{m1}^2 + y_{m2}} \tag{5.1}$$

$$y_{42} = y_{m1} + \sqrt{c - y_{m1}^2 + y_{m2}} \tag{5.2}$$

$$y_{43} = y_{m2} - \sqrt{c - y_{m2}^2 + y_{m1}} \tag{5.3}$$

$$y_{44} = y_{m2} + \sqrt{c - y_{m2}^2 + y_{m1}} \tag{5.4}$$

(2), (4)→
$$y_{m1} y_{m2} = -\frac{1}{4} \tag{6}$$

(1), (3), (6)→
$$y_{m1} = y_{m2}^2 + 4 y_{m1}^2 a^2 \varphi - c$$
$$y_{m2} = y_{m1}^2 + a^2 \varphi - c$$

Multiplying the upper/lower expression by $y_{m2}$/ $y_{m1}$ respectively, applying the relation (6) and addition of the two, we get successively:

$$y_{m1} y_{m2} = y_{m2}^3 + 4 y_{m1}^2 y_{m2} a^2 \varphi - c y_{m2}$$
$$y_{m1} y_{m2} = y_{m1}^3 + y_{m1} a^2 \varphi - c y_{m1}$$

$$-\frac{1}{4} = y_{m2}^3 - y_{m1} a^2 \varphi - c y_{m2}$$
$$-\frac{1}{4} = y_{m1}^3 + y_{m1} a^2 \varphi - c y_{m1}$$

$$y_{m1}^3 + y_{m2}^3 - c(y_{m1} + y_{m2}) + \frac{1}{2} = 0$$

Taking into account that $y_{m1} < 0$, and relation (6) we get successively:

$$y_{m1}^3 + y_{m2}^3 = y_{m2}^3 - |y_{m1}|^3 = (y_{m2} - |y_{m1}|)(y_{m1}^2 + y_{m2}^2 + y_{m2}|y_{m1}|)$$
$$y_{m1}^3 + y_{m2}^3 = (y_{m1} + y_{m2})(y_{m1}^2 + y_{m2}^2 - y_{m2} y_{m1}) =$$

$$= (y_{m1} + y_{m1})\left(y_{m1}^2 + y_{m2}^2 + 2 y_{m2} y_{m1} + \frac{3}{4}\right)$$

$$= (y_{m1} + y_{m2})\left((y_{m1} + y_{m2})^2 + \frac{3}{4}\right) = (y_{m1} + y_{m2})^3 + \frac{3}{4}(y_{m1} + y_{m2})$$

$$(y_{m1} + y_{m2})^3 + (y_{m1} + y_{m2})\left(\frac{3}{4} - c\right) + \frac{1}{2} = 0$$

Introducing term of fixed points functions mean value we finally get the equation:

$$Y_m = \frac{1}{2}(y_{m1} + y_{m2})$$

$$Y_m^3 + Y_m\left(\frac{3}{16} - \frac{c}{4}\right) + \frac{1}{16} = 0$$

Solving of this cubic polynomial equation yields real root:

$$Y_m = -\frac{1}{2}\left(\left(\frac{1}{4} - \sqrt{\frac{1}{16} - \left(-\frac{1}{4} + \frac{c}{3}\right)^3}\right)^{1/3} + \left(\frac{1}{4} + \sqrt{\frac{1}{16} - \left(-\frac{1}{4} + \frac{c}{3}\right)^3}\right)^{1/3}\right) \qquad (6)$$

which fully determines the exact analytical expression of fixed points functions.

The mean value functions for bifurcation branches are determined as follows:

$$y_{m1} + y_{m2} - 2 Y_m = 0$$

Multiplying by $y_{m1}$, applying (6) and solving the equation we get:

$$y_{m1}^2 - \frac{1}{4} - 2 Y_m y_{m1} = 0$$

$$y_{m1} = Y_m - \sqrt{\frac{1}{4} + Y_m^2} \qquad (7.1)$$

$$y_{m2} = Y_m + \sqrt{\frac{1}{4} + Y_m^2} \qquad (7.2)$$

Finally fixed points functions are determined as follows:

$$y_{41} = Y_m - \sqrt{\frac{1}{4} + Y_m^2} - \sqrt{c + Y_m + \sqrt{\frac{1}{4} + Y_m^2} - \left(Y_m - \sqrt{\frac{1}{4} + Y_m^2}\right)^2} \qquad (8.1)$$

$$y_{42} = Y_m - \sqrt{\frac{1}{4} + Y_m^2} + \sqrt{c + Y_m + \sqrt{\frac{1}{4} + Y_m^2} - \left(Y_m - \sqrt{\frac{1}{4} + Y_m^2}\right)^2}$$ (8.2)

$$y_{43} = Y_m + \sqrt{\frac{1}{4} + Y_m^2} - \sqrt{c + Y_m - \sqrt{\frac{1}{4} + Y_m^2} - \left(Y_m + \sqrt{\frac{1}{4} + Y_m^2}\right)^2}$$ (8.3)

$$y_{44} = Y_m + \sqrt{\frac{1}{4} + Y_m^2} + \sqrt{c + Y_m - \sqrt{\frac{1}{4} + Y_m^2} - \left(Y_m + \sqrt{\frac{1}{4} + Y_m^2}\right)^2}$$ (8.4)

Fully developed expressions using full expression of $Y_m$ as a function of $c$ is presented in the first part of the article, except that parameter $c$ has to change the sign from – to +. The same is valid for the expressions of period 1 and 2 fixed points functions and they are the following:

$$y_1 = \frac{1}{2}\left(1 - \sqrt{1 + 4c}\right)$$

$$y_{21} = \frac{1}{2}\left(-1 - \sqrt{-3 + 4c}\right)$$

$$y_{22} = \frac{1}{2}\left(-1 + \sqrt{-3 + 4c}\right)$$

Hereafter are shown the diagrams presenting the results as follows:

Diagram 1: Bifurcation diagram with plotted fixed points functions for cycles of period 1, 2 and 4.

Diagram 2: Presentation of fixed points functions of period 1, 2 and 4, mean value functions for period 4 and iterating functions from 1 to 8 ($y_1 = -c$).

Diagram 3: The same as in Diagram 2 but with the range of parameter limited from 0.75 to 1.575 and including iterating functions up to 16.

Diagram 4: The same as in Diagram 3 except that the range of parameter is further narrowed between 1.25 and 1.575 so as to make it clearly visible how the fixed points functions of period 4 pass through key intersection points of iterating functions 9 – 16, which in fact determine the space of bifurcation with cycle of period 8.

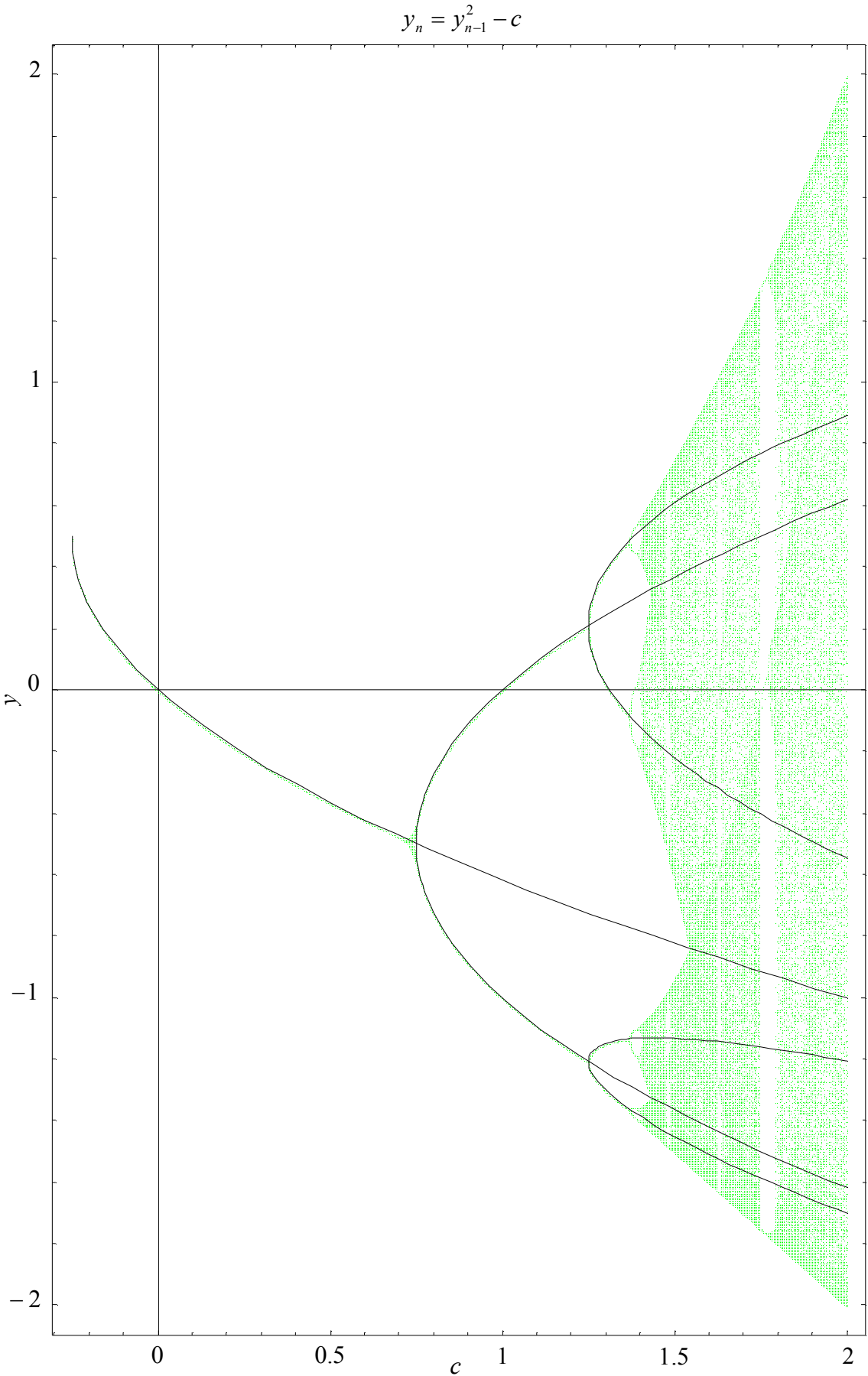

$$y_n = y_{n-1}^2 - c$$

Diagram 2

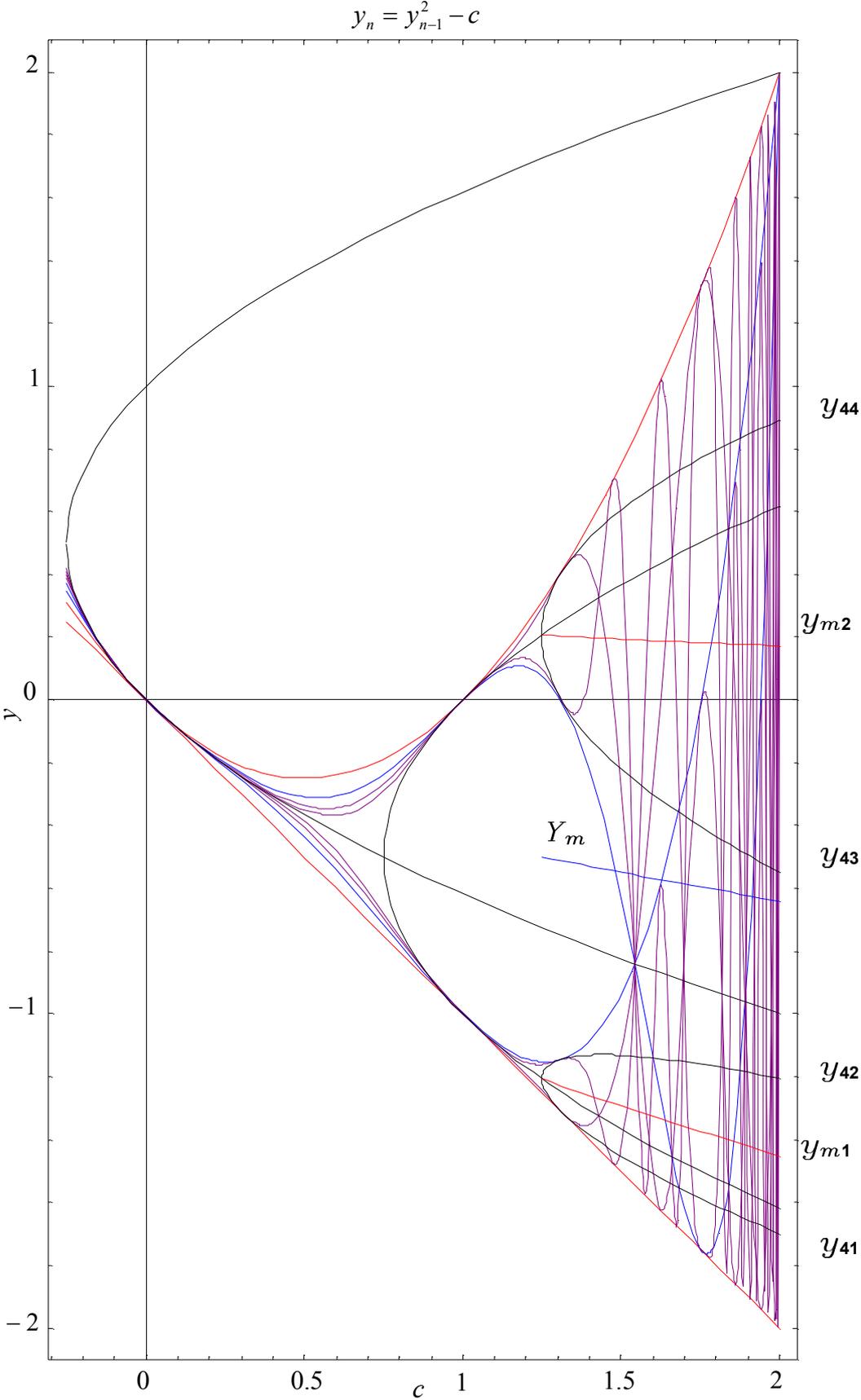

Diagram 3

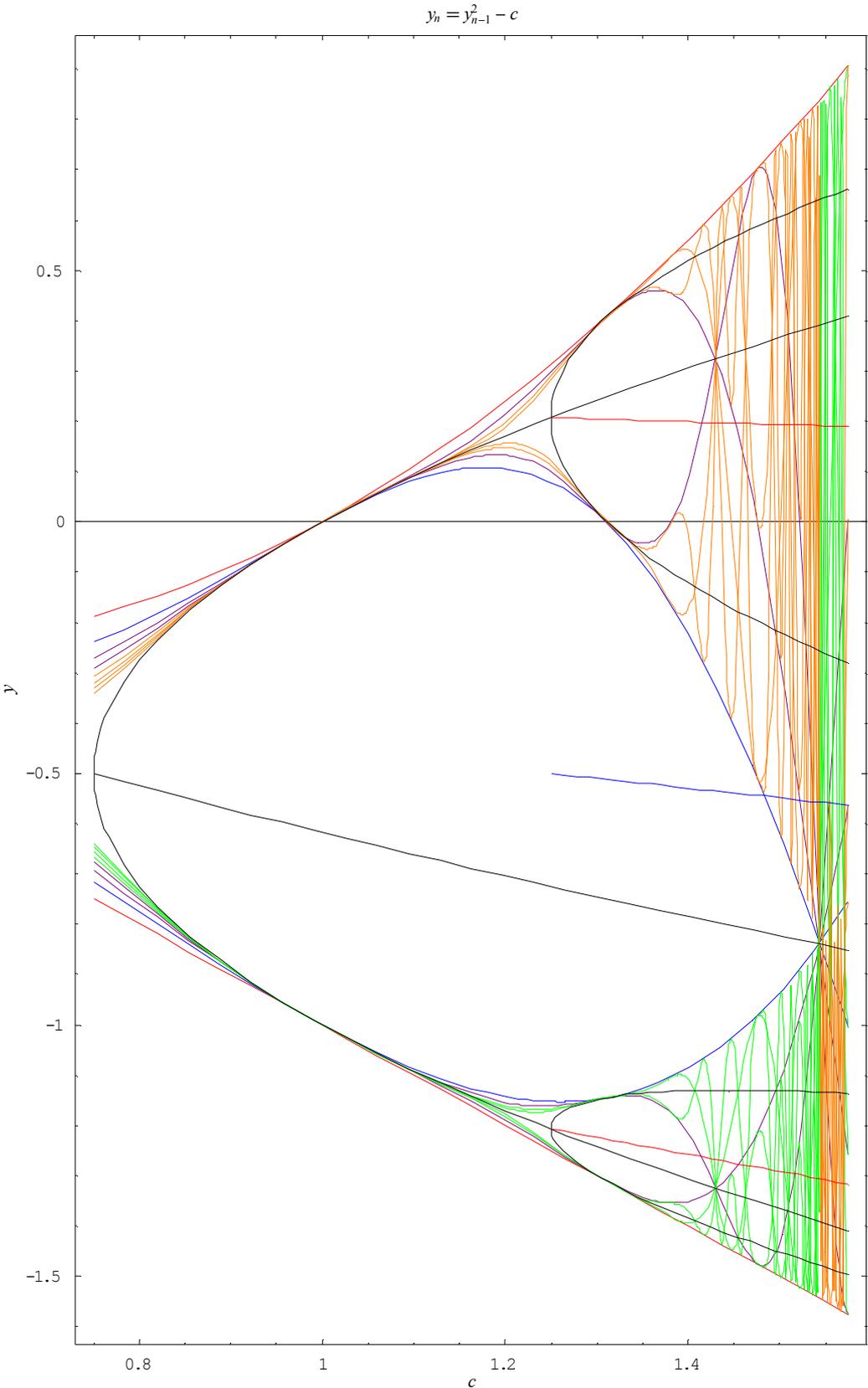

Diagram 4

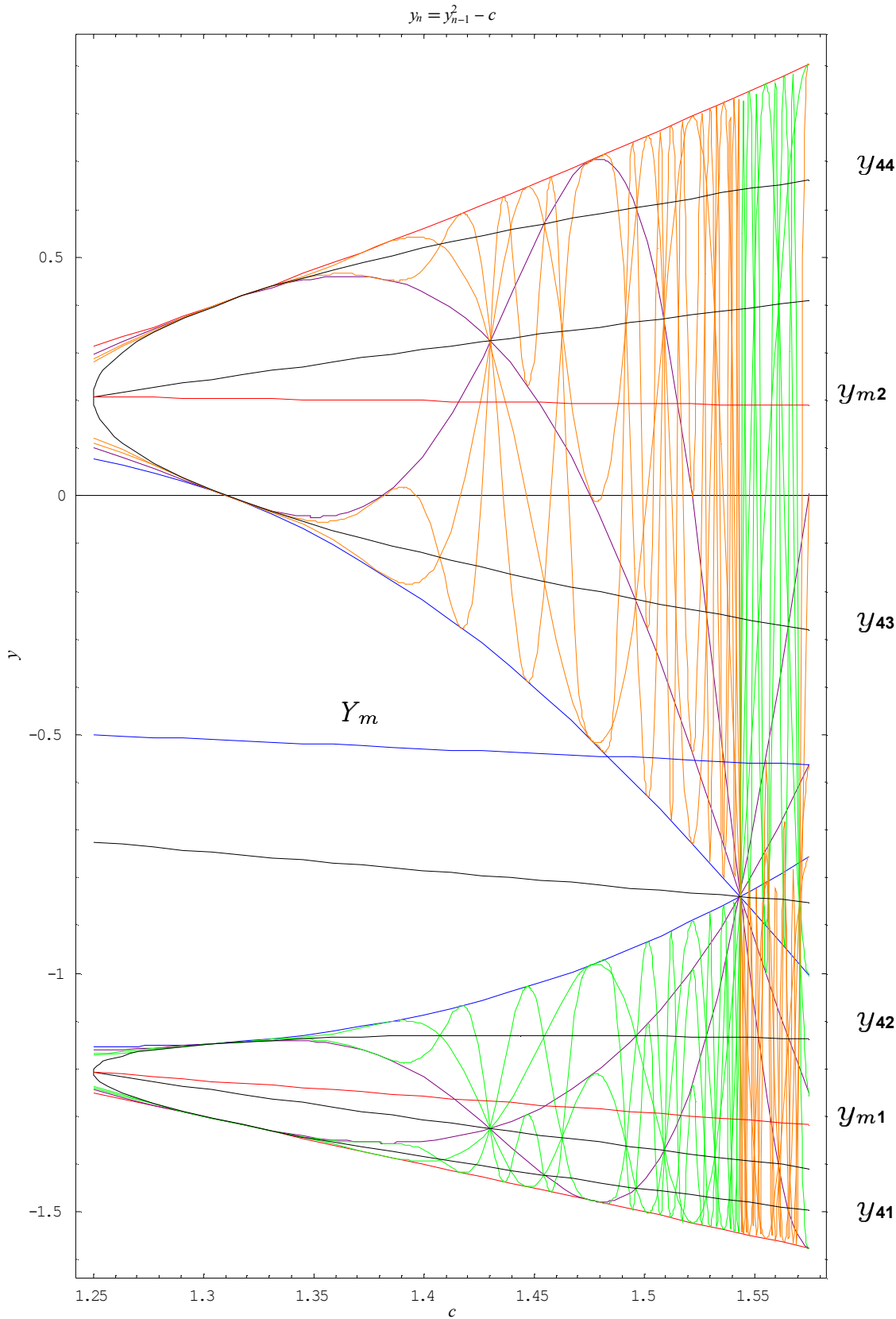

# 4. Conclusion

The presented solution of period 4 fixed points functions is in fact very simple and it offers the way for further investigations in the range of cycle of period 4 as well as research on higher period cycles. This approach, using mean values, is much more than the instrument for solving this particular case. It reveals many features of cycles with higher periods and I am convinced that the doors to further explorations and solutions are just opened.

Rijeka, 6 January 2008                                         Gvozden Rukavina

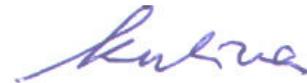

**References**

[1] Peitgen, H.-O.;Richter, P. H.: "The Beauty of Fractals", Springer-Verlag Berlin Heidelberg, 1986

[2] Peitgen, H.-O.; Jurgens, H.; Saupe, D.: "Fractals for the Classroom", Springer-Verlag New York, 1992